\documentclass[final,12pt,3p, times]{elsarticle}

\usepackage[utf8]{inputenc}
\usepackage[arabic, english]{babel}

\usepackage{comment}
\usepackage{lineno,hyperref}
\modulolinenumbers[5]
\usepackage{multirow}
\usepackage{makecell}
\usepackage{adjustbox}
\usepackage{xcolor}
\usepackage{booktabs}
\usepackage{titlesec}
\usepackage{enumitem}
\usepackage{lineno,hyperref}
\usepackage{url}
\usepackage{tabularx,diagbox,ragged2e}
\usepackage{adjustbox}
\usepackage{comment}
\usepackage{lineno,hyperref}
\usepackage{color}
\usepackage{graphics}
\usepackage{graphicx}
\usepackage{booktabs} 
\usepackage{longtable}
\usepackage{colortbl}
\usepackage{tikz}
\usepackage{tabularx}
\usepackage{makecell}
\usepackage{array}
\usepackage{balance}
\usepackage{framed}
\usepackage{amsmath}
\usepackage{blindtext}
\usepackage{enumitem}
\usepackage{multirow}
\usepackage{multicol}
\usepackage{empheq}
\usepackage{natbib}
\usepackage{subcaption}
\usepackage{caption}

\journal{arxiv.org}

\biboptions{authoryear}

\begin{document}
\begin{frontmatter}

\title{ ChatGPT and Beyond: The Generative AI Revolution in Education}
\author{Mohammad AL-Smadi\corref{cor1}}
\ead{malsmadi@qu.edu.qa}

\address{Qatar University, Qatar}

\cortext[cor1]{Corresponding author}

\begin{abstract}
The wide adoption and usage of generative artificial intelligence (AI) models, particularly ChatGPT, has sparked a surge in research exploring their potential applications in the educational landscape. This survey examines academic literature published between November, 2022, and July, 2023, specifically targeting high-impact research from Scopus-indexed Q1 and Q2 journals. This survey delves into the practical applications and implications of generative AI models across a diverse range of educational contexts. Through a comprehensive and rigorous evaluation of recent academic literature, this survey seeks to illuminate the evolving role of generative AI models, particularly ChatGPT, in education. By shedding light on the potential benefits, challenges, and emerging trends in this dynamic field, the survey endeavors to contribute to the understanding of the nexus between artificial intelligence and education. The findings of this review will empower educators, researchers, and policymakers to make informed decisions about the integration of AI technologies into learning environments.
  
\end{abstract}

\begin{keyword}
Artificial Intelligence \sep Education \sep ChatGPT \sep Generative AI \sep   Technology-enhanced Learning.
\end{keyword}
\end{frontmatter}

\section{Introduction} \label{Intro}

Generative artificial intelligence (AI) is developing rapidly with a great potential to revolutionize education. ChatGPT, a large language model (LLM), is a particularly promising generative AI model that can be applied in education. One of the key advantages of using generative AI in education is that it can help to personalize learning \citep{farrokhnia2023swot}. Generative AI models can be used to create personalized learning materials, such as practice problems, study guides, and feedback, tailored for each student \citep{doi:10.1080/14703297.2023.2190148}. This can help students learn in a way that is tailored to their individual needs and interests and at their own pace. Another advantage of using generative AI in education is that it can provide students with feedback and support outside of the classroom. For example, generative AI can be used to create chatbots that can answer students' questions and provide them with support \citep{doi:10.1080/08874417.2023.2261010}. Using Chatbots in education can be especially helpful for students who are struggling with learning or who need support outside of class time. Finally, generative AI can be used to automate tasks such as grading and creating reports. This can free up teachers' time to focus on more important tasks, such as interacting with students and providing them with personalized support.

While generative AI has the potential to revolutionize education, there are also some challenges that need to be addressed. One challenge is ensuring the accuracy and reliability of generated content \citep{glaser2023exploring}. Generative AI models can produce realistic but inaccurate content. It is important to develop mechanisms to ensure the accuracy and reliability of generated content, especially when it is used for educational purposes. Another challenge is addressing ethical concerns. The integration of generative AI in education raises ethical concerns and challenges, such as the potential for cheating and plagiarism, and teacher displacement \citep{DWIVEDI2023102642} . It is important to develop ethical guidelines for the use of generative AI in education and to educate students about these guidelines \citep{susarla2023janus}.

By addressing the challenges and conducting further research, we can ensure that generative AI is used to improve learning for all students. In this survey paper, we will explain the history and evolution of generative AI models and their application in education (Section \ref{History}). Moreover, we will review the current state of research on the use of generative AI and ChatGPT in education (Section \ref{sec_related}), and provide specific examples of how ChatGPT is being used in educational settings today (Section \ref{ChatGPT-education}). We will also discuss the potential benefits and challenges of using generative AI in education (Section \ref{Rel-advanteges}). We will conclude by discussing the survey findings and providing recommendations for future research (Section \ref{sec_discussion}, \ref{sec_Conclusion}).

\section{History of Using AI in Education} \label{History}
The history of using AI in education dates back to the 1960's, with the development of early intelligent tutoring systems. These systems were designed to provide personalized instruction to students, tailored to their individual needs and learning styles. However, before we delve into the evolution of using generative AI in education, we need to understand the history and evolution of generative AI models. 
\subsection{The History and Evolution of Generative AI Models}
\label{GAI-History}

Generative Artificial Intelligence (AI) models, particularly Language Models (LLMs), have witnessed remarkable progress over the years, transforming the landscape of natural language processing and a wide array of other creative tasks \citep{susarla2023janus}. In this section, we delve into the historical roots and evolutionary trajectory of these models, highlighting key milestones that have shaped their development.
\begin{itemize}
    \item \textbf{Early Days of Language Modeling:}
LLMs development history started in the 1950's and 1960's with the emergent of statistical Natural Language Processing (NLP). In its infancy, language models primarily employed statistical methodologies to estimate the likelihood of a given word or word sequence within a linguistic context. N-grams and sequences of n words were fundamental techniques during this period \citep{russell2010artificial}. 
\item \textbf{From N-grams to Word Embeddings:}
A pivotal shift from n-gram-based models to the use of word embeddings began to emerge in the mid-2000's with the introduction of the "Word2Vec" algorithm by \citep{mikolov2013distributed} in 2013. This innovative approach hinged on the utilization of vector representations to capture the semantic meaning of words. This breakthrough laid the groundwork for subsequent developments in language modeling.
\item \textbf{Advancements in Text-based deep learning models (i.e. Sequence-to-Sequence NLP):}
The integration of word embeddings into language modeling ushered in a new era. These vector representations served as input to deep learning models such as recurrent neural networks (RNNs) and, later, the encoder-decoder architecture. This shift had a profound impact on NLP research, including text summarization and machine translation, as demonstrated by \citep{sutskever2014sequence}. The ability to capture semantic context through vector representations significantly enhanced the quality and depth of generated content.
\item \textbf{The Transformer Architecture Revolution:}
The introduction of Transformer architecture by \citep{vaswani2017attention} in 2017 is considered as a turning point in the advancement of NLP and computer vision research and in particular in language modeling research. The transformer architecture represented a paradigm shift in NLP by introducing a self-attention mechanism.  Several deep learning models have been developed based on the transformer architecture such as BERT \citep{devlin2018bert}. This innovation enabled the model to capture long-range dependencies within sequences, improving the coherence and contextuality of generated content. The Transformer architecture laid the foundation for the subsequent development of LLMs.

\item \textbf{The Emergence of LLMs:}
In recent years, the field of AI witnessed the proliferation of Large Language Models (LLMs). These models which are also known by the term "foundation models" are trained on vast and diverse datasets encompassing books, news articles, web pages, and social media posts and tuned with billions of hyperparameters \citep{bommasani2021opportunities}. This unprecedented scale of data, coupled with advancements in model architecture and training techniques, marked a significant turning point. These foundation models exhibit an extraordinary adaptability to a wide range of tasks, including tasks for which they were not originally trained. ChatGPT stands as an exemplary case of a generative AI model in action. This remarkable AI system was launched in November 2022 and is fine-tuned from the generative pre-trained transformer GPT-3.5, which was originally trained on a large dataset of text and code sources \citep{neelakantan2022text}. ChatGPT harnesses the power of Reinforcement Learning from Human Feedback (RLHF), a technique that has shown immense promise in aligning Large Language Models (LLMs) with human intent \citep{christiano2017deep}. The astonishingly superior performance of ChatGPT underscores the potential for a paradigm shift in the training of generative AI models. This shift involves the adoption of instruction aligning techniques, such as reinforcement learning \citep{christiano2017deep}, prompt engineering \citep{brown2020language}, and chain-of-thought (CoT) prompts \citep{wei2022chain}, as a collective step toward the realization of building an ecosystem of intelligent services based on generative AI models.
\end{itemize}

The culmination of these advancements has led to generative AI models that possess a remarkable capacity to comprehend and generate media-rich realistic and proper content (including text, images, audio, and video). Such capabilities have enabled these models to be utilized and widely adopted in different applications such as education. Despite these advancements, concerns and challenges have arisen in the generative AI landscape \citep{susarla2023janus}. The ease with which models like ChatGPT can be adapted to new tasks raises questions about the depth of their understanding. Experts in AI fairness have warned against the potential for these models to perpetuate societal biases encoded in their training data \citep{glaser2023exploring}, labeling them as "stochastic parrots" \citep{bender2021dangers}. 

\subsection{Evolution of Using Generative AI in Education}
Using AI in Education is not new, the first attempts of using AI in Educations can be tracked back to the early 1960s, when researchers at the University of Illinois at Urbana-Champaign developed an intelligent tutoring system (ITS) called PLATO (Programmed Logic for Automatic Teaching Operations) \citep{bitzer1961plato}. PLATO was the first computer system that enabled students with graphical user interfaces to interact with educational materials that were developed and adapted using AI to their needs. Another example on early attempts of using AI in Education is the "Automatic Grader" system that was developed in the 1960's to automatically grade programming classes \citep{hollingsworth1960automatic}.

The advent of personal computers has increased the developments of ITSs during the 1970's, an example of a system that was developed in that period is TICCIT (Time-shared, Interactive Computer-Controlled Instructional Television) \citep{stetten1971ticcit}. TICCIT was another early ITS that was developed in the early 1970's at the University of Pittsburgh. TICCIT was an early attempt to deliver individualized multi-media based content in mass to users at homes and schools.   

The advancements in the developments of ITSs in 1960's and 1970's was backed up with learning theories and principles that value the one-to-one individualized tutoring of students at classrooms (See for example the work of B.F. Skinner's pioneering work on "programmed instruction movement" and Benjamin Bloom's work on "mastery learning"\citep{block1976mastery}. The developed ITSs during that period were mainly rule-based systems. Advancements in AI and the advent of micro-computers in the 1970's have influenced the way ITSs were trained and developed \citep{reiser2001history}. since the 1980's, the use of computer-based instruction and AI-based education in particular has evolved to automate several instructional activities \citep{reiser2001history2}.     

The arrival of the world-wide-web (WWW) in the 1990's has had a major shift in the delivery medium of intelligent educational services \cite{9069875}. ITSs have evolved to deliver intelligent, adaptive, and personalized learning services underpinned by machine learning models. Despite of these advancements in the way ITSs were developed and delivered to the users, their capabilities were limited to the delivery of individualized instruction and learning. The evolution of the WWW to the so called "Web 2.0" and the additional capabilities of collaborative and social based interaction has paved the way to a new era in the development of ITSs. Collected data based on users' interaction with the Web 2.0 services, and the ability of training software agents on these data using different machine learning algorithms has led to having more advancements in the application of learning analytics to adapt and personalized learning \citep{clow2013overview}.   

The 21st century has witnessed several breakthroughs in using AI in education. These breakthroughs were backed up by advancements in: (i) \textbf{Hardware capabilities and performance} \citep{nickolls2010gpu}, (ii) \textbf{Big data mining} \citep{wu2013data}, and (iii) \textbf{AI models and architectures} (i.e. the advent of deep learning models) \citep{lecun2015deep}. The advent of the Transformer deep learning architecture in 2017 \citep{vaswani2017attention}, is considered to be a turning point in the history of developing intelligent software in general (See Section \ref{GAI-History}). Many intelligent models such as generative pre-trained transformers (GPT) has started to appear right after \citep{radford2018improving}. In November 2022, OpenAI has released ChatGPT - which is based on GPT 3.5 architecture - and reached over 100 million users in just a few months. Since then,   and today generative AI-based educational tools are developed to provide students with personalized instruction, adaptive learning, and engaging learning experiences (See Section \ref{ChatGPT-education}).

\section{Research Methodology}
\label{sec_Methodology}
This section discusses the research methodology we followed to develop search, filter, and select the related work on using generative AI models in general and ChatGPT in particular in different educational settings.

\subsection{Search approach}
In this review, the Preferred Reporting Items for Systematic Reviews and Meta-Analysis (PRISMA) \citep{moher2009preferred,KNOBLOCH201191} statement was followed to select relevant articles. The search for articles started on June 25, 2023, until July 10. The version of ChatGPT discussed in the articles was the original release on November 30, 2022 (i.e. ChatGPT 3.5).  The search keywords ("ChatGPT", "Education", "Generative AI", "Chatpot") were used to search for related publications in different databases including ACM, IEEE Xplore, Scopus, and Web of Science. Articles that mentioned these keywords in their title, abstract, or keywords were selected. The publication period was filtered to be after 2022. Despite searching multiple databases, the number of relevant articles found was limited. To ensure a comprehensive search, Google Scholar was used to conduct a title search for the term "ChatGPT" within the same search period. This additional search method allowed for the identification of relevant articles that may not have been captured in the initial database search.

\subsection{Paper Selection Criteria}
In this review, academic articles published between November, 2022, and July, 2023, were included. The review included only advanced online publications published in journals indexed by Scopus as Q1 or Q2 journals. The focus of the review was on articles that discussed the application of generative AI models (mainly ChatGPT) in education, without any specific constraints on educational contexts. Identified literature reviews were used as background references and were excluded from the synthesis to avoid redundancy in references. Moreover, Only articles written in English were included in this rapid review, ensuring a consistent language focus for the analysis.

\section{Literature Review} \label{sec_related}

This section discusses the related research on using generative AI models and Chatpots in education with a focus on ChatGPT. Moreover, this section sheds the light on the potential advantages and challenges for using these technology in education.

\subsection{Chatbots in Education}\label{sec_related_english}

The increasing popularity of artificial intelligence (AI) chatbots in education has prompted numerous empirical studies to explore their impact on students' learning outcomes. However, the findings from these studies have been inconsistent, calling for a comprehensive review and synthesis. \cite{https://doi.org/10.1111/bjet.13334} is an example of a study that addresses this gap, where a meta-analysis of 24 randomized studies using Stata software (version 14) was conducted with an objective to analyze the impact of adopting AI chatbots on students' learning outcomes and to evaluate the moderating effects of educational levels and required intervention duration. The meta-analysis revealed that AI chatbots have a substantial effect on students' learning outcomes. Additionally, the effects of AI chatbots were found to be more pronounced among higher education students compared to those in primary and secondary education. Moreover, shorter interventions with AI chatbots were found to have a stronger effect on students' learning outcomes compared to longer interventions. This finding suggests that the novelty effects of AI chatbots may enhance learning outcomes in the short term but diminish over time in longer interventions. To further enhance students' learning outcomes, future designers and educators are encouraged to incorporate human-like avatars, gamification elements, and emotional intelligence into AI chatbot interventions. These features can potentially increase engagement and motivation, leading to improved learning outcomes.

\cite{https://doi.org/10.1111/bjet.13341} presented a comprehensive three-level model that integrates various learning theories to describe the functions of artificial intelligence (AI) in facilitating learning processes. This comprehensive model synthesizes multiple learning theories and offers valuable insights into the diverse roles AI can play in education. It explains how learning takes place at micro, meso, and macro levels. The research proposes fourteen distinct roles for AI in education, corresponding to the features of the model, and are different based on the application level ("four roles at the individual or micro level, four roles at the meso level of teams and knowledge communities, and six roles at the macro level of cultural historical activity"). By aligning AI functionalities with the model's structure, AI developers can collaborate with learning designers, researchers, and practitioners to enhance individual learning, team performance, and the establishment of knowledge communities. The model findings hold implications for the improvement of individual learning, team collaboration, and the cultivation of knowledge communities.


\cite{GUO2023104862} proposed Chatbot-assisted in-class debates (CaIcD) as a task design that integrates argumentative chatbots into classroom debates to enhance students' argumentation skills and task motivation. In CaIcD, students interact with an argumentative chatbot called Argumate before participating in debates with their peers. The chatbot helps students try ideas to support their position and prepare to opposing viewpoints. A study with 44 Chinese undergraduate students found that CaIcD led to significant improvements in students' argumentation skills. Students were able to generate arguments using more claims, data, and warrants. Their arguments were also more organized, sufficient, and elaborated. Additionally, students reported enjoying CaIcD more and exerting more effort than conventional learning tasks. Despite facing more challenges, students performed in CaIcD as successful as in conventional debates with their classmates.

In the same direction, \cite{IKUSILAN2023104812} investigated the effects of using "a decision-guided chatbot on interdisciplinary learning of learners with different cognitive styles, learning motivation, collective efficacy, classroom engagement, satisfaction with the learning approach, and cognitive load". The study was conducted with 71 junior high school students in northern Taiwan, who were divided into an experimental group (n=35) and a control group (n=36). The experimental group used a decision-guided chatbot for learning, while the control group used conventional technology-enhanced learning methods. The results showed that the experimental group using the decision-guided chatbot significantly outperformed the control group in terms research questions such as ("learning achievements, collective efficacy, extrinsic motivation, cognitive engagement, emotional engagement, and satisfaction with the learning approach"). Additionally, the experimental group reported lower mental effort during the learning process. Regarding cognitive styles, analytical learners achieved significantly higher learning achievements than intuitive learners, regardless of the learning method used. However, the study found that analytical learners in the experimental group experienced significantly lower mental load compared to intuitive learners.

\subsection{ChatGPT in Education} \label{ChatGPT-education}

ChatGPT, a generative language model developed by OpenAI, has gained significant popularity since its launch in November 2022, with over 100 million users in just a few months. The rapid advancement of generative AI-based tools is significantly impacting higher education institutions (HEIs), yet their influence on existing practices for continuous learning improvement has been underexplored, especially in undergraduate degree programs. With the wide usage of ChatGPT from students, it has become crucial for academic programs' designers to understand the implications of using AI-based tools on the results of evaluating students' achievement of the learning outcomes. These outcomes are commonly assessed to enhance program quality, teaching effectiveness, and overall learning support. To address this gap, this Section highlights related work on utilizing ChatGPT  - as a generative AI model - in several educational applications.

 \cite{doi:10.1080/2331186X.2023.2210461} conducted an empirical study to evaluate the capability of ChatGPT in solving a variety of assignments from different level courses in undergraduate degree programs. The performance of ChatGPT was then compared with the highest-scoring student(s) for each assignment and results were examined to understand their impact on the students' learning outcomes. Furthermore, the assignments generated by ChatGPT were assessed using widely used plagiarism detection tools such as Turnitin, GPTZero, and Copyleaks to determine their compliance with academic integrity standards. The findings emphasize the need for HEIs to adapt their practices to ensure academic integrity and monitor the quality of student learning. The study's results have important implications for HEI managers, regulators, and educators as they navigate the integration of AI tools in the educational landscape.

While academics have already utilized ChatGPT for tasks like drafting papers or conducting systematic reviews, its role in supporting learning has not been thoroughly examined.  To address this research gap, the author of \citep{stojanov2023learning} conducted a study reflecting on their personal experience of using ChatGPT as a "more knowledgeable other" to scaffold their learning on the technical aspects of how ChatGPT functions. The findings indicate that ChatGPT provided sufficient content to develop a general understanding of its technical aspects. The feedback received from ChatGPT was perceived as motivating and relevant. However, there were limitations observed during the interactions. The answers provided by ChatGPT were occasionally superficial, and the generated text was not always consistent, logical, or coherent. In some instances, contradictory responses were given. The immediate responses from ChatGPT contributed to a sense of flow and engagement during the learning process. However, this heightened engagement also led to an overestimation of knowledge and understanding, as the author was unable to immediately recognize contradictory information. Based on these observations, caution is advised when using ChatGPT as a learning aid. It is essential to consider the capabilities and limitations of the technology, as well as how humans perceive and interact with these AI systems. 

\cite{cooper2023examining} explored the potential of using ChatGPT to transform science education through investigating three main areas: (1) ChatGPT's ability to answer science education questions, (2) Ways for educators to use ChatGPT in their science pedagogy, and (3) How ChatGPT was used as a research tool in this study. The authors found that ChatGPT was able to generate comprehensive and informative responses to a wide range of science education questions, with its outputs often aligning with key themes in the research. Moreover, they suggests that educators can use ChatGPT to generate personalized learning materials, create interactive learning activities, provide students with feedback and support, and assist with the design of science units, rubrics, and quizzes. Nevertheless, the authors used ChatGPT to assist with editing and improving the clarity of their research narrative. They found that ChatGPT was a valuable tool for this purpose, but emphasize the importance of critically evaluating and adapting AI-generated content to the specific context. However, they also highlighted the ethical considerations associated with the use of AI in education, such as the risk of ChatGPT positioning itself as the ultimate authority and the potential for copyright infringement. They argue that it is important for educators to adopt responsible approach for using ChatGPT, prioritize critical thinking through focusing meta-cognitive skills in their provided learning activities, and clearly communicate expectations to their students.

\cite{su2023unlocking} investigated the potential benefits and challenges associated with using ChatGPT and generative AI in education. They introduced a theoretical framework, "IDEE," which guides the implementation of educative AI in educational settings. The framework includes four key components: "identifying desired outcomes, setting the appropriate level of automation, ensuring ethical considerations, and evaluating effectiveness". By exploring the opportunities and challenges within this framework, the framework contributes to the ongoing research on the integration of AI technologies in education.

\cite{tlili2023if} conducted a three-stage study where in the in the first stage, the researchers investigated the community sentiment of using ChatGPT on different social media platforms. The overall sentiment was found to be positive, with enthusiasm for its potential use in educational settings. However, there are also voices of caution regarding its implementation in education. In the second stage of the study, they analyzed the case of using ChatGPT in education from multiple perspectives, including "educational transformation, response quality, usefulness, personality and emotion, and ethics". Based on the stage findings, the authors provided insights into the various dimensions and considerations associated with integrating ChatGPT into educational contexts. During the final stage, ten educational scenarios were implemented to investigate the user experiences of using ChatGPT in these scenarios. Several issues were identified, including concerns about cheating, the honesty and truthfulness of ChatGPT's responses, privacy implications, and the potential for manipulation. Overall, this study sheds light on the use of ChatGPT in education, providing insights into public discourse, different dimensions of its implementation, and user experiences. The findings emphasize the importance of considering ethical implications and ensuring responsible adoption of chatbots in education.

\cite{10105236} conducted a two-stage study aimed to investigate how senior students in a computer engineering program perceive and evaluate ChatGPT's impact on teaching and learning. The study included 56 participants, and in the first stage, students were asked to provide their own evaluations of ChatGPT after completing a learning activity using the tool. The researchers analyzed the students' responses (3136 words) through coding and theme building, resulting in 36 codes and 15 themes. In the second stage, three weeks later, the students had to complete a questionnaire of 27 items after being engaged in other activities with the assistance of ChatGPT. The findings indicated that students agreed on the the interesting capabilities of ChatGPT, finding it easy-to-used, motivating, and supportive for studying. They appreciated its ability to  provide well-structured answers and explanations. However, students also expressed concerns about the accuracy of ChatGPT's responses and the need  for having background knowledge to evaluate them. Students opinions were also divided on the negative impact of ChatGPT on learning performance, academic integrity and dishonesty, and job displacement. The study concludes that ChatGPT can and should be used to augment teaching and learning, both teachers and students must build competencies on how to properly use it, developers are urged to enhance the accuracy of the answers provided by ChatGPT.


\cite{doi:10.1080/10494820.2023.2209881} developed a model to examine the main predictors of adoption and use of ChatGPT in higher education. 534 student at a Polish state university participated in this study and their self-reported data were used to evaluate the research hypotheses. Out of the ten hypotheses proposed, nine were confirmed by the results. Findings revealed that "habit" was the most reliable indicator of "behavioral intention", followed by "performance expectancy" and "hedonic motivation". When it came to actual use, "behavioral intention" was the most significant factor, followed by "personal innovativeness". The research also highlights the need for continued exploration and understanding of how AI tools can be effectively integrated into educational settings.


Technology-enhanced language learning, particularly the use of generative AI models, have positively engaged and enhanced the performance of second language (L2) learners. With the introduction of ChatGPT, educators have utilized its automatic text generation capabilities in writing classrooms. For instance, \cite{yan2023impact} explored the application of ChatGPT's text generation capabilities in a one-week L2 writing activity and examine students' behaviors and reflections during their usage of ChatGPT. The findings demonstrated the potential applicability of ChatGPT in L2 writing pedagogy, highlighting its affordances as a tool for composing writing efficiently. However, participants also expressed concerns about the tool's implications for academic honesty and educational equity. Specifically, there were worries about the potential for plagiarism and the need to redefine plagiarism in the context of generative AI technology. This prompted a call for the development of regulatory policies and pedagogical guidance to ensure the proper utilization of ChatGPT in educational settings.

Gunther Eysenbach, founder and publisher of JMIR Publications, conducted an interview with ChatGPT to explore the potential of using chatbots in medical education \citep{eysenbach2023role}. The interview offers a glimpse into the current capabilities and challenges of ChatGPT, showcasing its potential to support medical education while also emphasizing the need for careful editing, critical evaluation, and human oversight in utilizing AI technology in the medical field. While the language model occasionally makes mistakes, it acknowledges and corrects them when challenged. However, it also demonstrates a tendency to generate fictional references, highlighting the known issue of large language models sometimes producing inaccurate or fabricated information. As a result, JMIR Medical Education is launching a call for papers on AI-supported medical education, with the initial draft of the call for papers being generated by ChatGPT and subsequently edited by human guest editors.

\cite{mcgee2023using} employed ChatGPT to conduct a search for citations related to a specific book. While ChatGPT did not provide the exact list of citations requested, it offered valuable information on how to compile such a list. Despite not yielding the intended outcome, the exercise was not considered useless because ChatGPT provided guidance and insights that could be useful in the process of generating the desired list of citations. This suggests that ChatGPT, even if it did not directly fulfill the initial request, still offered valuable assistance and potential for future research endeavors.

\cite{foroughi2023determinants} reported an interesting study, where 406 Malaysian students participated in a study to identify possible determinants of intention to use ChatGPT in education. The study builds on the Unified Theory of Acceptance and Use of Technology 2 (UTAUT2) and collected data was analyzed using a hybrid approach that combines "partial least squares" (PLS) and "fuzzy-set qualitative comparative analysis" (fsQCA). Based on the PLS analysis factors such as performance expectancy, effort expectancy, hedonic motivation, and learning value were found to be significant determinants of intention to use ChatGPT. More precisely, findings suggest that students are more likely to use ChatGPT if they believe that it will help them to perform better, learn more easily, and have a more enjoyable learning experience. Interestingly, the PLS analysis shows that factors like social influence, facilitating conditions, and habit do not have a significant impact on ChatGPT use. However, the fsQCA analysis suggests that these factors may indeed affect the intention to use ChatGPT, indicating that there might be different combinations of factors that lead to high ChatGPT use.

\cite{smith2023old} explored the potential use of ChatGPT to support teaching methods in the of social psychiatry. The researchers interacted with ChatGPT 3.5 to perform one of the tasks related to their course. Based on their experiences, the researchers found that ChatGPT can be used effectively to teach and learn active and case-based learning activities for both students and instructors in social psychiatry. However, they acknowledge the limitations of chatbots, such as the potential for misinformation and biases. They suggest that these limitations may be temporary as the technology continues to advance. The researchers argue that with appropriate caution, generative AI models can support social psychiatry education, and they encourage further research in this area to better understand their potential and limitations.

\cite{choi2023chatgpt}, sought to assess the ability of the generative AI model (i.e. ChatGPT) to generate answers for law school exams without human assistance. They used ChatGPT to generate answers for four real exams at the University of Minnesota Law School. The exams consisted of over 95 multiple choice questions and 12 essay questions. The researchers then blindly graded the exams as part of their regular grading processes. The results indicated that ChatGPT performed at the level of a C+ student on average across all four courses. While the performance was relatively low, ChatGPT was able to achieve a passing grade in all four exams. The study provides detailed information about these results and discusses their implications for legal education. Furthermore, the study offers guidelines on how ChatGPT can assist with legal writing including example prompts and advice.

\cite{mohammed2023psychometric} reported the development, validation, and utilization of a tool called the Knowledge, Attitude, and Practice (KAP-C). KAP-C was developed with an aim of using ChatGPT in pharmacy practice and education. More precisely, how pharmacists and pharmacy students perceived and behaved while using ChatGPT in their professional practice and educational settings. The authors explored the literature related to using ChatGPT in supporting Knowledge, Attitude, and Practice in several educational setting focusing on health related studies. Results from literature review were used to develop a survey targeting  "pharmacists and pharmacy students in selected low- and middle-income countries (LMICs) including Nigeria, Pakistan, and Yemen". Findings from this survey will provide insights into the psychometric properties of the KAP-C tool and assess the knowledge, attitude, and practice of pharmacy professionals and students in LMICs towards using ChatGPT.

\cite{fergus2023evaluating} investigated the potential impact of using ChatGPT in learning and assessment. The study focuses on two chemistry modules in a pharmaceutical science program, specifically examining ChatGPT-generated responses in answering end-of-year exams. The study finds that ChatGPT is capable of generating responses for questions that require knowledge and understanding, indicated by verbs such as "describe" and "discuss." However, it demonstrated a limited level in answering questions that involve knowledge application and interpretation of non-text information. Importantly, findings of the study suggests that ChatGPT is not a high-risk technology tool for cheating. However, it emphasizes that the use of ChatGPT in education, similar to the disruptions caused by the COVID-19 pandemic, stimulates discussions on assessment design and academic integrity.

This editorial \citep{Lodge_Thompson_Corrin_2023} highlights the profound impact of generative artificial intelligence (AI) on tertiary education and identifies key areas that require rethinking and further research in the short to medium term. The rapid advancements in large language models and associated applications have created a transformative environment that demands attention and exploration. The editorial acknowledges that while there has been prior research on AI in education, the current landscape is unprecedented, and the AI in education community may not have fully anticipated these developments. The editorial also outlines the position of AJET (Australasian Journal of Educational Technology) regarding generative AI, particularly in relation to authors using tools like ChatGPT in their research or writing processes. The evolving nature of this field is acknowledged, and the editorial aims to provide some clarity while recognizing the need to revisit and update the information as the field continues to evolve in the coming weeks and months.


\subsection{Advantages and Challenges for Using Generative AI Models in Education} \label{Rel-advanteges}

Generative AI models have the potential to revolutionize education by personalizing learning, providing feedback and support, and automating tasks. However, there are also some challenges that need to be addressed before they can be widely adopted in the classroom. The utilization of generative AI models in academia has become a prominent and debated topic within the education field. For instance, ChatGPT, offers numerous advantages, such as enhanced students' learning and  engagement, collaboration, and learning accessibility. However, its implementation also raises concerns regarding academic integrity and plagiarism \citep{susarla2023janus}. This Section sheds the light on related work from literature that address these challenges and paves the way forward to a proper adoption of Generative AI models in Education.

 \cite{doi:10.1080/14703297.2023.2190148} explored the opportunities and challenges associated with the use of ChatGPT in higher education. One significant challenge is the detection and prevention of academic dishonesty facilitated by generative AI models. The paper highlights the difficulties in identifying plagiarism when AI-generated content is involved. To address these concerns, the paper proposes strategies that HEIs can adopt to ensure responsible and ethical utilization of generative AI models like ChatGPT. These strategies include the development of policies and procedures that explicitly outline the expectations and guidelines for using AI tools. Additionally, the study emphasizes on providing training and support to both students and faculty members as crucial step towards promoting an understanding and responsible usage of these technologies. Employing various methods to detect and prevent cheating, such as plagiarism detection software and proctoring technologies, is also recommended. The authors also argue that by implementing strategies that prioritize academic integrity, universities can harness the benefits of generative AI models while mitigating potential risks.

\cite{chan2023comprehensive} investigated the perceptions and implications of text generative AI technologies in higher education institutions in Hong Kong. The study involved 457 students and 180 teachers and staff from various disciplines. Accordingly, the study proposes an "AI Ecological Education Policy Framework" with an aim of addressing the versatile implications of integrating AI technologies into the educational system at the university. The framework is organized into three dimensions: "Pedagogical, Governance, and Operational". The Pedagogical dimension of the framework emphasizes the use of AI to enhance teaching and learning outcomes. It explores how AI can be leveraged to improve instructional practices, personalize learning experiences, and facilitate student engagement. The Governance dimension of the framework focuses on addressing issues related to privacy, security, and accountability in the context of AI integration. It emphasizes the establishment of policies, guidelines, and safeguards to protect data privacy, ensure ethical AI practices, and promote responsible use of AI technologies. The Operational dimension of the framework deals with practical considerations such as infrastructure and training. It highlights the importance of providing the necessary technological infrastructure and support for successful AI implementation in higher education. Additionally, it emphasizes the need for training programs to equip educators and staff with the competences required to effectively use AI technologies.

\cite{KASNECI2023102274} argue that a responsible approach of adopting LLMs in education must be preceded by enabling teachers and learners with specific set of competencies and training to master the use of generative AI tools, its limitations, and the potential vulnerabilities of such systems. They emphasize the need for clear strategies and pedagogical approaches that prioritize critical thinking and fact-checking skills to fully integrate large language models into learning settings and teaching curricula. The commentary provides recommendations for addressing these challenges and ensuring the responsible and ethical use of large language models in education. The authors ague that by addressing these concerns and adopting a sensible approach, large language models can be effectively utilized to enhance educational experiences while educating students about the potential biases and risks of AI applications.

 \citep{adiguzel2023revolutionizing} discussed the advantages and challenges associated with the use of generative AI models in education, aiming to provide valuable insights for incorporating these technologies responsibly and ethically in educational settings. They also highlighted the advantages of implementing generative AI models in education. Mainly, improved learning outcomes, increased productivity, and enhanced student engagement through personalized education, feedback, and assistance. However, the authors acknowledge the ethical and practical challenges that arise when implementing AI in education. Key concerns include potential bias in AI algorithms and the need for adequate teacher/student preparation and support. In their study, the authors emphasized the potential applications of generative AI models, specifically chatbots and ChatGPT, in areas such as personalized learning, language instruction, and feedback provision. 

\cite{mohamed2023exploring} conducted a survey study to explore the perceptions on using ChatGPT in supporting students' English language learning. Ten English as a Foreign Language (EFL) faculty members at Northern Border University participated in this study. The study employed in-depth interviews as the primary method of data collection. The findings from the interviews revealed a range of opinions among the faculty members regarding the efficacy of using ChatGPT. Some participants acknowledged the usefulness of implementing ChatGPT in answering quickly and accurately a wide array of questions. They recognized its potential to complement and enhance traditional EFL teaching methods. On the other hand, some faculty members expressed concerns that the use of ChatGPT may hinder students' development of critical thinking and research skills. They also highlighted the potential for ChatGPT to reinforce biases or propagate misinformation. Overall, the faculty members viewed ChatGPT as a valuable tool for supporting EFL teaching and learning. However, they recognized the need for further research to assess its effectiveness. 

\cite{TSAI202371} conducted a study with a focuses on the advantages of incorporating generative AI models into chemical engineering education. For this purpose, the researchers proposed an approach that utilizes ChatGPT as an learning aided tool to solve problems in the field of chemical engineering. The research study aimed at addressing the limitation in chemical engineering education and deepen students' understanding of core related subjects through focusing on practical problem-solving instead of theoretical knowledge. The researchers conducted an experimental lecture where they presented a simple example of using ChatGPT to "calculate steam turbine cycle efficiency". They also assigned projects to students to explore the potential applications of ChatGPT in solving problems in chemical engineering. The collected feedback from students was mixed, but overall, ChatGPT was reported to be an accessible and practical tool for improving their abilities in problem-solving. The researchers also highlighted the problem of Hallucination where generative AI models can generate realistic but inaccurate content, which could be misleading and negatively affects the learning process.

\cite{wardat2023chatgpt} reported their experience on using ChatGPT for teaching mathematics based on the results of a two-stage study to evaluate both content generated by ChatGPT and the perceptions of the users. The authors reported a limitation on ChatGPT capabilities in understanding geometry problems with a variation based on the the complexity of the equations, input data, and instructions given to ChatGPT. However, it is anticipated that the capabilities of ChatGPT will become more efficient in solving geometry problems in the future. Regarding the perceptions of the users, the public discourse on social media reflects increased interest for using ChatGPT in teaching mathematics. However, there are also voices of caution regarding its use in education. The findings of this study recommends further research to ensure the safe and responsible use of chatbots, particularly ChatGPT, into mathematics education and learning. 

Despite the rapid increase in adopting generative AI models and ChatGPT in particular in several educational scenarios and applications, there is still limited research on how educators should properly use this technology to augment their instruction. For instance,
the Authors of this study \citep{jeon2023large}  investigated how teachers and ChatGPT, a large language model chatbot, can work together to improve education. Eleven language teachers used ChatGPT in their classes for two weeks. After that, they shared their experiences in interviews and provided logs of their interactions with ChatGPT. The analysis of the data showed that ChatGPT could play four roles in education: "interlocutor (having conversations with students), content provider (providing information), teaching assistant (helping students with their work), and evaluator (assessing student learning)". These roles showed how ChatGPT could be used in different parts of the teaching process. Additionally, the data showed that teachers played three roles: "(a) orchestrating different resources with good pedagogical decisions, (b) making students active learners, and (c) raising awareness of AI ethics". These roles highlighted the importance of teachers' teaching expertise in using AI tools effectively. The findings showed that teachers and AI can work together effectively to improve education. Teachers need to use their teaching expertise to incorporate AI tools into their instruction. The study provided insights into how large language model chatbots could be used in education in the future and why it is important for teachers and AI to work together.


\cite{choudhury2023investigating} investigated the impact of users' trust in ChatGPT on their intentions and actual usage of the technology. The study examined four hypotheses concerning the connection between trust, the intention to use, and the actual usage of ChatGPT. The authors conducted a web-based survey targeting adults in the United States who actively engage with ChatGPT (version 3.5) at least once a month. The responses from this survey were employed to construct two latent variables: Trust and Intention to Use, while Actual Usage served as the dependent variable. A total of 607 participants completed the survey, with primary ChatGPT use cases revolving around information acquisition, entertainment, and problem-solving. The results from the structural equation modeling revealed that trust exerts a substantial direct influence on both the intention to use ChatGPT and its actual usage. Results demonstrated a 50.5\% of the variation in the intention to use and 9.8\% of the variation in actual usage. Through bootstrapping, the authors validated all four null hypotheses, affirming the significant impact of trust on both intention and actual usage. Furthermore, they identified that trust indirectly affects actual usage, partially mediated by intention to use. These findings underscore the pivotal role of trust in the adoption of ChatGPT. Nevertheless, the authors also argues that, it's essential to acknowledge that ChatGPT wasn't explicitly designed for healthcare applications, and overreliance on using it for health-related advice can lead to misinformation and potential health hazards. The study stresses on the necessity of enhancing ChatGPT's capacity to distinguish between queries it can properly answer and those that should be handled by human experts. Moreover, the study recommends shared responsibility and collaboration among developers, subject matter experts, and human factors researchers to mitigate the risks linked to an excessive trust in generative AI models such as ChatGPT.

\cite{farrokhnia2023swot} conducted a SWOT analysis of ChatGPT to examine its strengths, weaknesses, opportunities, and threats in the context of education. The strengths of ChatGPT identified in the analysis include its ability to generate plausible answers using a sophisticated natural language model, its capacity for self-improvement, and its potential to provide personalized and real-time responses. These strengths suggest that ChatGPT can enhance access to information, facilitate personalized and complex learning experiences, and reduce the workload for educators, thereby improving efficiency in educational processes. However, the analysis also highlights several weaknesses of ChatGPT, such as its lack of deep understanding, lack of the ability to evaluating the quality of its responses, the potential for bias and discrimination, and its limited ability to foster higher-order thinking skills. These weaknesses pose challenges to its effective use in educational contexts. The study further discusses the opportunities that ChatGPT presents for education, including the potential for enhanced student engagement, individualized instruction, and support for diverse learners. It also explores the threats that ChatGPT poses to education, such as the risk of misunderstanding context, compromising academic integrity, perpetuating discrimination in educational practices, enabling plagiarism, and potentially diminishing higher-order cognitive skills.

\label{sec_background}

\section{Summary}
\label{sec_discussion}

\begin{table}[!t]
\centering
\begin{tabular}{|p{5cm}|p{10cm}|}
\hline
\textbf{Advantage} & \textbf{Description} \\
\hline
Personalized Learning & Generative AI models can create personalized learning materials, such as practice problems, study guides, and feedback, tailored to each student's needs and interests \citep{adiguzel2023revolutionizing, farrokhnia2023swot}. This can help students learn at their own pace and in a way that is tailored to their individual needs and interests. \\
\hline
Feedback and Support & Generative AI models can be used to provide students with feedback and support outside of the classroom \citep{mohamed2023exploring}. For example, generative AI models can be used to create chatbots that can answer students' questions and provide them with support. \\
\hline
Automate Tasks & Generative AI models can be used to automate tasks such as grading and creating reports \citep{jeon2023large}. This can give teachers more time to focus on other important tasks, such as interacting with students and providing them with personalized support. \\
\hline
Enhanced Student Engagement & Generative AI models can be used to enhance students' engagement by providing them more personalised learning scenarios, timely feedback, and opportunities to interact and create unique content \citep{farrokhnia2023swot}, such as poems, stories, and code \citep{TSAI202371, jeon2023large} \\
\hline
\end{tabular}
\caption{Advantages of Using Generative AI Models in Education}
\label{tab:advantages-education}
\end{table}

\begin{table}[!t]
\centering
\begin{tabular}{|p{5cm}|p{10cm}|}
\hline
\textbf{Challenge} & \textbf{Description} \\
\hline
Ensuring Accuracy and Reliability  (Hallucination) & Generative AI models can generate realistic but inaccurate content, potentially causing the spread of misinformation \citep{farrokhnia2023swot,stojanov2023learning, eysenbach2023role, TSAI202371}. Mechanisms are needed to ensure accuracy and reliability of generated content, especially in sensitive and educational contexts \citep{yan2023impact}. \\
\hline
Addressing academic dishonest & The use of generative AI models in education raises ethical concerns, such as the potential for cheating and plagiarism \citep{farrokhnia2023swot}.  Generative AI models can be used to generate essays, assignments, or answers, enabling academic dishonesty and plagiarism \citep{doi:10.1080/14703297.2023.2190148}. Educational institutions should take measures to prevent and detect academic dishonest, and invest in technologies and practices for detecting and preventing cheating facilitated by generative AI models. Ethical guidelines should be developed for their use in education, and students need to be educated about these guidelines \citep{tlili2023if, cooper2023examining}. Moreover,  It is essential to educate students on the ethical use of generative AI tools and reinforce the importance of academic integrity \citep{chan2023comprehensive}.
\\
\hline
Equity and Access & Equitable access to generative AI models is essential, and these models should be designed to be inclusive and accessible to all students \citep{yan2023impact, farrokhnia2023swot}. \\
\hline
Training and Support & Teachers and students need training on how to use generative AI models effectively in the classroom \citep{KASNECI2023102274}. They also require support in developing and implementing innovative learning activities that leverage the potential of generative AI models. \citep{10105236, doi:10.1080/14703297.2023.2190148, chan2023comprehensive} \\
\hline
Limitations in different disciplines & The limitation of generative AI capabilities in understanding and solving problems in some disciplines (such as geometry \citep{wardat2023chatgpt}, health-related advice \citep{choudhury2023investigating}) causes these models to generate inaccurate or wrong answers (i.e. Hallucination problem).\\
\hline
\end{tabular}
\caption{Challenges of Using Generative AI Models in Education}
\label{tab:challenges-education}
\end{table}

\begin{table}[!t]
\centering
\begin{tabular}{|p{5cm}|p{10cm}|}
\hline
\textbf{Ethical Consideration} & \textbf{Description} \\
\hline
Bias & Generative AI models can inherit biases from training data, leading to biased output \citep{smith2023old, KASNECI2023102274, mohamed2023exploring}. Awareness and mitigation of bias in AI-generated content are essential. \\
\hline
Privacy & Generative AI models may use data containing sensitive personal information, necessitating privacy protection for individuals \citep{chan2023comprehensive}. \\
\hline
Autonomy & As AI models become more autonomous, ethical implications arise. Safeguards should be developed to prevent models from making harmful decisions. \\
\hline
Job displacement & Generative AI models could automate tasks that are currently performed by human teachers and other education professionals. This could lead to job displacement and unemployment \citep{DWIVEDI2023102642, 10105236}.\\
\hline
\end{tabular}
\caption{Ethical Considerations of Using Generative AI Models}
\label{tab:ethical-considerations}
\end{table}

The integration of generative AI models in education offer a wide range of potential benefits, including improved productivity, enhanced creativity, and personalized learning, providing timely feedback and support, and automating tasks \citep{farrokhnia2023swot, doi:10.1080/14703297.2023.2190148}. The advantages of using generative AI models in education are summarised in Table \ref{tab:advantages-education}. However, the adoption of generative AI models in education raises challenges that need to be tackled in order to guarantee a safe and responsible usage of such technologies \citep{su2023unlocking}. These challenges are summarised in Table \ref{tab:challenges-education}.

\begin{table}[!t]
\centering
\begin{tabular}{|p{5cm}|p{10cm}|}
\hline
\textbf{Legal Implication} & \textbf{Description} \\
\hline
Copyright Infringement & Generative AI models can be used to generate copyrighted content without permission, raising copyright implications. Steps should be taken to avoid copyright infringement. \\
\hline
Intellectual Property Ownership & Ownership of intellectual property rights to content generated by generative AI models is unclear, leading to potential disputes over ownership and licensing. \\
\hline
Liability & Uncertainty regarding liability for harm caused by AI-generated content can pose legal challenges for businesses and organizations using generative AI models. \\
\hline
\end{tabular}
\caption{Legal Implications of Using Generative AI Models}
\label{tab:legal-implications}
\end{table}

\begin{table}[!t]
\centering
\begin{tabular}{|p{5cm}|p{10cm}|}
\hline
\textbf{Stakeholder} & \textbf{Recommendations} \\
\hline
Models developers & 
Focus on developing features that enhance performance expectancy, effort expectancy, hedonic motivation, and learning value for students \citep{foroughi2023determinants}. Also, consider ways to make generative AI models more appealing to students who are more innovative and who are more critical of the information they consume. Develop plugins to integrate with different learning management systems that supports different services such as, automated content generation, automated assessment feedback, and learning analytics. Consider adding human-like avatars, gamification elements, and emotional intelligence to these interventions \citep{https://doi.org/10.1111/bjet.13334}. These features can potentially make students more engaged and motivated, which could lead to better learning outcomes. Integrate explainable AI methods to make AI models more transparent and reliable.\\
\hline
Instructors & 
Provide students with opportunities to use generative AI models in ways that are aligned with their learning goals. Also, teach students how to critically evaluate the information they receive from generative AI models \citep{jeon2023large}. Redesign your assessment practices to focus more on meta-cognitive skills and higher level of the knowledge domain where generative AI models have limited capabilities to act\citep{fergus2023evaluating}. \citep{cooper2023examining}\\
\hline
Universities & 
Support the development of training and resources to help instructors and students use generative AI models effectively. Also, investigate ways to integrate generative AI models into the curriculum and to assess student learning outcomes associated with generative AI models use. Develop policies and guidelines that maintains a safe and responsible usage of generative AI in teaching and learning \citep{doi:10.1080/2331186X.2023.2210461, yan2023impact, doi:10.1080/14703297.2023.2190148}.\\
\hline
\end{tabular}
\caption{Insights Recommended for for Different Stakeholders to Properly Adopt Generative AI Models in Education.}
\label{tab:Recommendations}
\end{table}

Findings of this survey underscore the need for further research, regulatory policies, and pedagogical guidance to navigate issues related to academic integrity and equity in the era of AI-enhanced learning  \citep{doi:10.1080/2331186X.2023.2210461, yan2023impact, doi:10.1080/14703297.2023.2190148}. The irresponsible implementation of generative AI models in education have ethical and legal consequences \citep{doi:10.1080/2331186X.2023.2210461}. Table \ref{tab:ethical-considerations} discusses potential ethical consideration for the adoption of Generative AI models in education. Whereas, Table \ref{tab:legal-implications} focuses on the legal implication. 

Governments around the world are developing regulations to address the ethical and legal challenges posed by generative AI models. For example, the "European Union Artificial Intelligence Act (EU AI Act)" was proposed to promote the development and use of trustworthy AI, while also mitigating the risks associated with AI \citep{madiega2021artificial}. The Act does this by establishing a risk-based approach to AI regulation. AI systems are classified into four risk categories: "unacceptable risk, high risk, limited risk, and minimal risk". The EU AI Act includes the following provisions: (a) \textbf{Risk Assessment:} the AI Act mandates the assessment of risks associated with AI systems, ensuring that potential harms are identified and mitigated, (b) \textbf{Transparency:} transparency requirements are specified to enhance the explainability of AI systems and make their decision-making processes more understandable,  the EU AI Act specified the following three transparency requirements when it comes to adopt generative AI technology:

\begin{itemize}
  \item \textbf{Disclosure of AI Generation:} The imperative for transparency in the context of generative AI pertains to the acknowledgment and notification that content has been produced by artificial intelligence rather than a human agent. Such disclosures serve to elucidate the nature of the content creation process, fostering user understanding and trust.

  \item \textbf{Mitigation of Illicit Content Generation:} The design and implementation of generative AI models must encompass measures to forestall the generation of content that contravenes legal statutes. These measures are integral to ethical and legal compliance, involving content moderation and safety mechanisms to curtail the production of potentially harmful, offensive, or unlawful materials.

  \item \textbf{Publication of Summaries of Copyrighted Data Utilized in Training:} In the sphere of AI model development, the utilization of copyrighted data engenders concerns pertaining to intellectual property rights. To address these concerns and exemplify transparency, developers may be encouraged to provide summaries or descriptions of the copyrighted data sources employed during training. This step showcases adherence to copyright laws while enhancing transparency regarding the model's training data.
\end{itemize}

 and (c) \textbf{Human Oversight:} the AI Act emphasizes the importance of human oversight in AI systems to ensure responsible and ethical use. However, there are some limitations to the current regulatory landscape \citep{schuett2023risk}. For example, the EU AI Act does not explicitly address the issue of generative AI models hallucination. Additionally, the EU AI Act focuses on the obligations of AI developers and users, rather than on the rights of individuals who may be affected by AI systems \citep{enqvist2023human}.

Findings from literature also provide valuable insights into the determinants of intention to use generative AI models for educational purposes. The findings have important implications for the models developers, instructors, and universities as they work to accelerate the adoption of them in educational contexts. Table \ref{tab:Recommendations} discusses a set of recommendations for the different stakeholders fostering them to have a safe and responsible adoption of such technologies in education. 

The adoption or generative AI models in education requires a trust raising plan \citep{choudhury2023investigating}. In addition to addressing the forementioned challenges (See Table \ref{tab:challenges-education}), there are a number of steps that can be done to build trust in generative AI, including: (1) \textbf{Transparency:} developers should be transparent about the limitations, biases, data sources, and potential risks of these models. (2) \textbf{Explainability:} developers should develop explainable AI methods to explain how their models generate outputs \citep{dovsilovic2018explainable}. (3) \textbf{Human-in-the-loop validation:} human experts should be involved in the evaluation of AI-generated content to identify and rectify errors. This will help into solving other problems such as bias, and hallucination. (4) \textbf{Accountability:} developers and users of generative AI should be accountable for the consequences of their actions.

\section{Conclusion}
\label{sec_Conclusion}

Using ChatGPT and other generative AI tools in education offers several benefits. It allows for a more personalized and efficient learning experience for students, as the technology can adapt to individual needs and provide tailored support. Additionally, it enables teachers to deliver feedback more quickly and easily, enhancing the learning process. ChatGPT plays several roles in education, including providing information, facilitating debates and discussions, supporting self-directed learning, and creating content for course materials. In response to a specific prompt, ChatGPT can generate cases for learning specific topics. However, there are also challenges to consider. The effectiveness of the technology in educational settings is still largely untested, and there may be limitations in the quality of data that AI chatbots rely on. Ethical considerations, such as privacy and bias, as well as safety concerns, must also be addressed when implementing ChatGPT or similar tools in education. By addressing the challenges posed by AI technologies and leveraging their advantages, a fair and effective education system that provides individualized teaching, feedback, and support can be built.

This survey sheds light on the relationship between ChatGPT and teachers, revealing the different roles that each entity can play in the educational context. It emphasizes the importance of teachers' adapted pedagogical expertise while using such technology and highlights the potential usage of generative AI models to enhance instructional practices.

As a pioneering effort, this survey emphasized the need for future research to provide deeper insights into the application of generative AI models in teaching and learning. It also emphasized the importance of making appropriate pedagogical adjustments to effectively integrate these models into instruction. Moreover, this study highlights the need for a collaboration among educators, researchers, and policy-makers to develop regulatory guidelines and practices that ensure the ethical and responsible use of generative AI models in education.


\bibliographystyle{elsarticle-num}


\end{document}